\documentclass[12pt]{article}
\title{\centerline \bf Gauge transformation through an
  accelerated frame of reference} 
\bigskip 
\author{Ashish Shukla, Kaushik
  Bhattacharya \thanks{e-mail addresses:ashuklas@iitk.ac.in, 
kaushikb@iitk.ac.in,
  }\\
  \normalsize
  Department of Physics,\\
  \normalsize
  Indian Institute of Technology, Kanpur\\
  \normalsize
  Kanpur 208016, Uttar Pradesh, India.\\
}
\begin{document}
%%%%%%%%%%%%%%%%%%%%%%%%%%%%%%%%%%%%%%%%%%%%%%%
\maketitle
%%%%%%%%%%%%%%%%%%%%%%%%%%%%%%%%%%%%%%%%%%%%%%%
\begin{abstract}
The Schr\"{o}dinger equation of a charged particle in a uniform
electric field can be specified in either a time-independent or a
time-dependent gauge. The wave-function solutions in these two gauges
are related by a phase-factor reflecting the gauge symmetry of the
problem. In this article we show that the effect of such a gauge
transformation connecting the two wave-functions can be mimicked by
the effect of two successive extended Galilean transformations
connecting the two wave-function. An extended Galilean transformation
connects two reference frames out of which one is accelerating with
respect to the other.
\end{abstract}
%%%%%%%%%%%%%%%%%%%%%%%%%%%%%%%%%%%%%%%%%%%%%%%
\section{Introduction}
%%%%%%%%%%%%%%%%%%%%%
The Schr\"{o}dinger equation of a charged particle in presence of a uniform
electric field can be described in different gauges. Out of these various
gauges a uniform electric field can be described by a time-independent scalar
potential or time-dependent vector potential.  The solutions of the
Schr\"{o}dinger equation in these two gauges will be linked to each other by a
phase-factor. In this article we will focus on the relationship of the
solutions of the Schr\"{o}dinger equation in different gauges with the
solution of the same equation in a specific accelerated frame of reference.
This specific frame has the same rectilinear uniform acceleration as felt by a
classical charged particle in the electric field. The coordinates of the
accelerated frame and the inertial coordinates in which the original
Schr\"{o}dinger equation is written are related by an extended Galilean
transformation (EGT).  An EGT is simply a coordinate transformation which
connects two mutually accelerating coordinate systems. It is observed that the
non-inertial frame acts like a bridge between two different descriptions of
the Schr\"{o}dinger equation in two different gauges. Starting from any one
gauge in the inertial coordinates we can make an EGT to reach the non-inertial
coordinates and then again make a reverse EGT to come back to the inertial
coordinates. The point of interest is that, through this loop of EGTs, we may
come back to the Schr\"{o}dinger equation described in a different gauge in
the inertial coordinates.  Consequently two EGTs act like a gauge
transformation.  This is the central result of this article. We will also
briefly address the issue related to the existence of a strange non-stationary
solution of the Schr\"{o}dinger equation, with a time-independent potential,
which naturally arises for the case of the charged particle in an electric
field, first noted by Buch and Denman in Ref.~\cite{denman}.

The material of the article is presented in the following way. The next
section sets the notation and terminology used in this article.  Section
\ref{qac} discusses how two different solutions of the Schr\"{o}dinger
equation written in two different gauges can be connected via an accelerated
frame of reference. In section \ref{rel} the unusual solutions of the 
Schr\"{o}dinger equation in various cases will be discussed. The
unusual solutions turn out to be offshoots of general time-dependent
coordinate transformations. The concluding section \ref{concl} will end with a
brief comment on the results obtained in this article.
%%%%%%%%%%%%%%%%%%%%%%%%%%%%%%%%%%%%%%%%%%%%%%%%%%%%%%%%%%%%%%%%%%%
\section{Gauge  transformations and some preliminary comments}
%%%%%%%%%%%%%%%%%%%%%%%%%%%%%%%%%%%%
In non-relativistic quantum mechanics the Schr\"{o}dinger equation for
a particle in presence of a classical electrostatic field, in one
dimension, is given by
\begin{eqnarray}
\left[\frac{1}{2m}\left(-i\hbar\frac{\partial}{\partial x}-
\frac{q A}{c}\right)^2 + q\phi\right]\psi=i\hbar\frac{\partial
\psi}{\partial t}\,,
\label{sch}
\end{eqnarray}
where 
$$E=-\frac{\partial \phi}{\partial x} - 
\frac1c\frac{\partial A}{\partial t}\,.$$ The uniform electric field
along the $x$-axis can be specified via two gauges:
\begin{eqnarray}
\phi = -E_0 x\,\,,\,\,A_x=0\,,
\label{g1}
\end{eqnarray}
and
\begin{eqnarray}
\phi = 0\,\,,\,\,A_x=- E_0 c t\,,
\label{g2}
\end{eqnarray}
where $E_0$ is the magnitude of the electric field. For future
reference we will call the two gauges as the static and the dynamic
gauges. The solutions of the Schr\"{o}dinger equation in the above two
gauges are not unique. Unless there exists a uniqueness theorem for
solutions, a second order partial differential equation can have
multiple solutions. For an illustrative example, two possible
solutions of the Schr\"{o}dinger equation in the static gauge are
given below. The first one is \cite{landau}
\begin{eqnarray}
\psi_1(x,t) = \exp{\left(-\frac{i\epsilon t}{\hbar}\right)}
{\rm Ai(-\zeta)}
\label{stata}
\end{eqnarray} 
where ${\rm Ai}$ is the Airy function of the first order,
$\epsilon$ is a constant and
$$\zeta=\left(\frac{2m}{q^2 \hbar^2 E_0^2}\right)^{1/3}(\epsilon + q E_0
x)\,. $$ The second solution is \cite{denman}
\begin{eqnarray}
\psi_2(x,t) = \exp\left[-it\left(\frac{p^2 + q^2E_0^2t^2/3 - 2q
E_0xm}{2m\hbar}\right)\right]\exp\left[\pm\frac{ip}{\hbar}\left(
x- \frac{qE_0t^2}{2m}\right)\right]\,,
\label{statb}
\end{eqnarray}
where $p$ is a constant. In a similar way two possible solutions of
the Schr\"{o}dinger equation in the dynamic gauge are as follows:
\begin{eqnarray}
{\tilde\psi}_1(x,t) = \exp\left[-\frac{it}{\hbar}(\epsilon + qE_0x)
\right]{\rm Ai}(-\zeta) \,,
\label{dyna}
\end{eqnarray} 
and
\begin{eqnarray} 
{\tilde\psi}_2(x,t) = \exp\left[-it\left(\frac{p^2 + q^2E_0^2t^2/3}
{2m\hbar}
\right)\right]\exp\left[\pm\frac{ip}{\hbar}\left(x-
\frac{qE_0t^2}{2m}\right)\right]\,.
\label{dynb}
\end{eqnarray}
As the static gauge and the dynamic gauge are related by a gauge
transformation, the wave-functions $\psi_i(x,t)$ and ${\tilde\psi}_i(x,t)$,
where $i=1,\,2$, are related by a phase-factor. A one-to-one correspondence
exists between the solutions of the Schr\"{o}dinger equation due to gauge
symmetry. 

Before we proceed to the next section it is worth pointing out that the
solution of the Schr\"{o}dinger equation named $\psi_2(x,t)$ is a very strange
solution to accept, it gives a non-stationary solution to the Schr\"{o}dinger
equation in a time-independent potential. Mathematically it is a valid
solution of the differential equation in the static gauge and it also appears
to arise when we start from ${\tilde\psi}_2(x,t)$ in the dynamic gauge and
then make a gauge transformation to the static gauge. We will discuss about
this strange solution in section \ref{rel} where it will be shown that
$\psi_2(x,t)$ arises as the result of a time-dependent coordinate
transformation acting on the free particle wave-function.
%%%%%%%%%%%%%%%%%%%%%%%%%%%%%%%%%%%%%%%%%%%%%%%%%%%%%%%%%%%%%%%
\section{Equivalence of gauge transformation with two extended 
Galilean transformations}
\label{qac}
%%%%%%%%%%%%%%%%%%%%%%%%%%%%%%%%%%%%%%%%%%%%%%%%%%%%%
The appropriate Schr\"{o}dinger equation in the static gauge is
\begin{eqnarray}
-\frac{\hbar^2}{2m}\frac{\partial^2 \psi(x,t)}{\partial x^2}
- qE_0x \psi(x,t)=i\hbar\frac{\partial \psi(x,t)}{\partial t}\,.
\label{sch1in}
\end{eqnarray}
In this section we will suppress the index $i$ attached to the wave-functions
with the general understanding that all the wave-functions appearing in any
equation has the same subscript. The last equation is separable and one of the
solutions of the Schr\"{o}dinger equation is given in Eq.~(\ref{stata}). If we
make an EGT specified by
\begin{eqnarray}
\xi = x - \eta(t)\,,\,\,\,\,\,\,\tau = t\,,\,
\,\,(\dot{\eta}(t)\ll c)
\label{noninc}
\end{eqnarray}
the Schr\"{o}dinger equation corresponding to Eq.~(\ref{sch1in}) becomes
\begin{eqnarray}
-\frac{\hbar^2}{2m}\frac{\partial^2 \psi^{\prime\prime}(\xi,\tau)}{\partial 
\xi^2} -\left[ qE_0(\xi + \eta) - m \xi \ddot{\eta}\right]
\psi^{\prime\prime}(\xi,\tau)=i\hbar\frac{\partial 
\psi^{\prime\prime}(\xi, \tau)}{\partial \tau}\,.
\label{intm1}
\end{eqnarray}
Here $\eta(t)$ can be any function of $t$ specifying a rectilinear
displacement. The acceleration of the moving coordinate system $(\xi,\tau)$ is
$\ddot{\eta}$. In a nonrelativistic framework the velocity $\dot{\eta}$ is
always less than the velocity of light. The wave-functions $\psi(x,t)$ and 
$\psi^{\prime\prime}(\xi, \tau)$ are related by \cite{rosen, holstein, paddy}:
\begin{eqnarray}
\psi^{\prime\prime}(\xi,\tau)=\exp\left[-\frac{i}{\hbar}f(\xi,\tau)\right]
\psi(x,t)\,,
\label{coor_gauge1} 
\end{eqnarray}
where $\psi(x,t)$ on the right hand side of the above equation means
$\psi(\xi+\eta, \tau)$ and 
\begin{eqnarray}
f(\xi, \tau)=\int_0^\tau
\frac12 m\dot{\eta}^2\,dt + m\xi \dot{\eta}(\tau)\,.
\label{f}
\end{eqnarray}
For a specific choice of $\eta(t)=\frac{qE_0t^2}{2m}$, it is seen that 
Eq.~(\ref{intm1}) transforms into a free-particle equation: 
\begin{eqnarray}
-\frac{\hbar^2}{2m}\frac{\partial^2 \psi^\prime(\xi,\tau)}{\partial \xi^2}
 =i\hbar\frac{\partial 
\psi^{\prime}(\xi, \tau)}{\partial \tau}\,,
\label{freenin}
\end{eqnarray}
in the $(\xi, \tau)$ coordinates for 
\begin{eqnarray}
\psi^{\prime}(\xi, \tau)&=&\exp\left[-\frac{iq^2E_0^2\tau^3}{6m\hbar}\right]
\psi^{\prime\prime}(\xi,\tau)\nonumber\\
&=&\exp\left[-\frac{i}{\hbar}\left\{f(\xi,\tau) + \frac{q^2E_0^2\tau^3}{6m}
\right\}\right]\psi(x,t)\,,
\label{coor_gauge}
\end{eqnarray}
where Eq.~(\ref{coor_gauge1}) has been used to write the final form of the
above equation.

Starting from the free-particle equation as given in Eq.~(\ref{freenin}) in
the non-inertial coordinates, if we go back to $(x,t)$ coordinates then
obviously we can get back Eq.~(\ref{sch1in}). But Eq.~(\ref{sch1in}) is not
the unique equation which we get when we transform Eq.~(\ref{freenin}) into
the $(x,t)$ coordinates. Using coordinate transformations in
Eq.~(\ref{noninc}) the free-particle equation in $(\xi, \tau)$ coordinates
transform into
\begin{eqnarray}
-\frac{\hbar^2}{2m}\frac{\partial^2 \psi^\prime(x,t)}{\partial x^2} -
i \hbar \dot{\eta}\frac{\partial \psi^\prime(x,t)}{\partial x} 
=i\hbar \frac{\partial \psi^\prime(x,t)}{\partial t}\,, 
\label{schdyn1}
\end{eqnarray}
where $\psi^\prime(x,t)$ is equivalent to $\psi^\prime(\xi,\tau)$
expressed in terms of $(x,t)$. The form of the Schr\"{o}dinger
equation in the dynamic gauge is
\begin{eqnarray}
-\frac{\hbar^2}{2m}\frac{\partial^2 {\tilde \psi}(x,t)}{\partial x^2} -
i \hbar \dot{\eta}\frac{\partial {\tilde \psi}(x,t)}{\partial x} 
+ \frac{(qE_0t)^2}{2m}{\tilde \psi(x,t)}
=i\hbar \frac{\partial {\tilde \psi}(x,t)}{\partial t}\,, 
\label{schdyn2}
\end{eqnarray}
where $\eta=qE_0t^2/2m$. From the  form of the above equations it 
can be easily shown that,
\begin{eqnarray}
{\tilde \psi}(x,t)=\exp\left(-\frac{iq^2 E_0^2 t^3}{6m\hbar}\right)
\psi^\prime(\xi,\tau)\,.
\label{impr}
\end{eqnarray}
Eq.~(\ref{coor_gauge}) and Eq.~(\ref{impr}) combined together
establishes our result. Combining the two equations it is seen that,
if one starts in the static gauge using the inertial coordinates
$(x,t)$ and then apply an extended Galilean transformation with an
acceleration $\ddot{\eta}$ then 
$\psi(x,t)\,\,{\stackrel{\ddot{\eta}}\longrightarrow}\,\,\psi^\prime
(\xi,\tau)$. Under a second extended Galilean
transformation from the $(\xi,\tau)$ coordinates back to $(x,t)$ with
an acceleration $-{\ddot\eta}$ we have 
$\psi^\prime(\xi,\tau)\,\,{\stackrel{-\ddot{\eta}}\longrightarrow}\,\,
{\tilde\psi(x,t)}$. Consequently the gauge transformation $\psi(x,t)
\,\,{\stackrel{\rm gt}\longrightarrow}\,\,{\tilde \psi(x,t)}$ can be 
thought of as being produced by two extended Galilean transformations.
%%%%%%%%%%%%%%%%%%%%%%%%%%%%%%%%%%%%%%%%%%%%%%%%%%%%%%%%%%%%%%%%%%%%%%%
\section{The unusual solutions and extended Galilean relativity}
\label{rel}
%%%%%%%%%%%%%%%%%%%%%%%%%%%%%%%%%%%%%%%%%%%%%
If we choose $\psi_1(x,t)$ as the solution of the Schr\"{o}dinger equation in
the static gauge as given in Eq.~(\ref{stata}) and then use
Eq.~(\ref{coor_gauge}) to find the corresponding wave-function
$\psi^\prime_1(\xi,\tau)$ in the accelerated frame, where the particle is free,
we get:
\begin{eqnarray}
\psi^\prime_1(\xi,\tau)= \exp\left[\frac{-i\tau}{\hbar}\left(
\frac{q^2 E_0^2 \tau^2}{3m} + qE_0\xi + \epsilon
\right)\right]\,{\rm Ai}\left(-\left(\frac{2mqE_0}{\hbar^2}\right)^{1/3}
\left(\xi + \frac{qE_0\tau^2}{2m} + \frac{\epsilon}{qE_0}\right)\right)\,.
\nonumber
\end{eqnarray}
As $\psi^\prime_1(\xi,\tau)$ is the solution of 
Eq.~(\ref{freenin}) we can replace $\xi$ by $\xi^\prime$ where
$\xi^\prime=-\xi - (\epsilon/qE_0)$ which amounts to changing the coordinate
origin to $-(\epsilon/qE_0)$ and then changing the sign of $\xi$. In terms of 
$\xi^\prime$ we can write
\begin{eqnarray}
\psi^\prime_1(\xi^\prime,\tau)= \exp\left[\frac{iqE_0\tau}{\hbar}\left(
\xi^\prime - \frac{q^2 E_0^2 \tau^2}{3m}
\right)\right]\,{\rm Ai}\left(\left(\frac{2mqE_0}{\hbar^2}\right)^{1/3}
\left(\xi^\prime - \frac{qE_0\tau^2}{2m}\right)\right)\,.
\label{berry}
\end{eqnarray}
The above solution is an accelerating solution of the Schr\"{o}dinger
equation, the acceleration given by $qE_0/m$, first observed by Berry and
Balazs \cite{berry}. Berry and Balazs attributed the existence of the
accelerating Airy function solution to the existence of a family of particles
each of which are moving with a uniform velocity.  But there exists an
alternative interpretation of the accelerating solution.  From the very
existence of the accelerating solution of the free particle speaks about the
existence of an accelerated frame where the free particle does not remain
free anymore, a point previously noted by \cite{greenberger}. From the
principle of extended Galilean relativity we cannot eliminate the accelerated
frame as a valid frame of reference. The wave-functions in the inertial and
non-inertial frames will then be connected by a phase relationship giving rise
to a solution like the one appearing in Eq.~(\ref{berry}).

The free-particle equation in $(\xi,\tau)$ coordinates, Eq.~(\ref{freenin}),
also has the usual plane-wave solution given by
\begin{eqnarray}
\psi_2^\prime(\xi,\tau)=
\exp\left[\frac{i}{\hbar}(\pm p\xi - E_p
  \tau)\right]\,,\,\,\,\,\,\,E_p=\frac{p^2}{2m}\,.
\end{eqnarray}
where $p$ stands for the momentum eigenvalue in $(\xi,\tau)$
coordinates.  To an observer in $(\xi,\tau)$ coordinates the
coordinate system $(x,t)$ seems to be accelerating. The free-particle
solution can be now transformed to the $(x,t)$ coordinates by
\begin{eqnarray}
\psi_2(x,t)=\exp\left[-\frac{i}{\hbar}f(x,t)\right]
\psi_2^\prime(\xi,\tau)\,,
\label{coor_gauge2} 
\end{eqnarray}
where 
\begin{eqnarray}
f(x, t)=\int_0^t \frac12
m\dot{\eta}^2\,d\tau - mx\dot{\eta}(t) = \frac{q^2E_0^2t^3}{6m} - xqE_0t\,.
\label{fxt}
\end{eqnarray}
Then the corresponding wave function in the $(x,t)$ coordinates is
$$\psi_2(x,t)=\exp\left[-\frac{it}{\hbar}\left(\frac{q^2E_0^2t^2}{6m} -
  xqE_0\right)\right]\exp\left[\frac{i}{\hbar}\left\{\pm p\left(x -
  \frac{qE_0t^2}{2m}\right) - E_p t\right\}\right]\,,$$ which is equal
to the solution of the Schr\"{o}dinger equation in the static gauge as
given in Eq.~(\ref{statb}). 

The plane-wave solution of a charged particle in presence of an electric field
and the accelerating solution of the free-particle both defies common sense.
But, their existence is a necessity as, without them the description of the
problem will not remain relativistic in the extended Galilean sense.  
%%%%%%%%%%%%%%%%%%%%%%%%%%%%%%%%%%%%%%%%%%%%%%%%%%%%%%%%%%%%%%%%%%%%%%%%%%
\section{Conclusion}
\label{concl}
%%%%%%%%%%%%%%%%%%%%%%%%%%%%%%%%%%%%
In this article we observe that there is an arbitrariness of the form of the
Schr\"{o}dinger equation under two EGTs with equal and opposite accelerations
connecting two mutually accelerating reference frames. This arbitrariness
arises from the fact that two Lagrangians that differ from each other by the
total time derivative of a function actually represents the same physical
situation.  An arbitrariness also appears in the form of the Schr\"{o}dinger
equation describing the properties of a charged particle in an electric field,
as the differential equation can be represented in two different gauges. It
turns out that these two very different forms of arbitrariness of the problem
are interlinked with each other. This is the reason why the phase-factor
connecting the two wave-functions in different gauges can be reproduced by the
effect of two successive EGTs connecting the two wave-functions. 

In this article we also address the issue concerning the time-dependent
solution of a time-independent Schr\"{o}dinger equation in the static gauge as
manifested in $\psi_2(x,t)$. It is shown that $\psi_2(x,t)$ can be thought of
as arising from the effect of an EGT acting on a free-particle wave-function
written in an accelerated coordinate system, or, as a result of a gauge
transformation acting on ${\tilde\psi}_2(x,t)$. This observation shows that the
dynamic gauge description of the problem has a close analogy with the
free-particle description of the problem. The exact relationship between these
two descriptions is represented in Eq.~(\ref{impr}).

It is interesting to note that there exists a one-to-one correspondence of the
Schr\"{o}dinger equation of a charged particle in presence of an electrostatic
field as given in Eq.~(\ref{sch1in}) with the Schr\"{o}dinger equation of a
massive particle in presence of a constant gravitational force $mg$. Using
this similarity we can find out the wave-function of the massive gravitating
particle in the accelerated frame using exactly the same techniques used in
section \ref{qac}. But, the similarity of the two situations ends there as,
there is no corresponding gauge principle for Newtonian gravity and
consequently there is no equation which corresponds to Eq.~(\ref{schdyn2}) for
the gravitational case.
%%%%%%%%%%%%%%%%%%%%%%%%%%%%%%%%%%%%%%%%%%%%%%%%%%%%%%%%%%%%%%%%

%%%%%%%%%%%%%%%%%%%%%%%%%%
\end{document}